%% file: angular_dep-main_arxiv_v2.tex
\documentclass[fleqn,10pt]{wlscirep}
\usepackage{amsmath}
\usepackage{nicefrac}
\usepackage[resetlabels,labeled]{multibib}
\newcites{S}{References}

\title{Angular dependence of vortex instability in a layered superconductor: the case study of Fe(Se,Te) material}

\author[1,*,+]{Gaia Grimaldi}
\author[2,1,+]{Antonio Leo}
\author[2,1]{Angela Nigro}
\author[2,1]{Sandro Pace}
\author[3]{Valeria Braccini}
\author[3]{Emilio Bellingeri}
\author[3]{Carlo Ferdeghini}
\affil[1]{CNR SPIN, Salerno, Fisciano, 84084, Italy}
\affil[2]{University of Salerno, Physics Department, Fisciano, 84084, Italy}
\affil[3]{CNR SPIN, Genova, 16152, Italy}
\affil[*]{gaia.grimaldi@spin.cnr.it}
\affil[+]{these authors contributed equally to this work}

\begin{abstract}

Anisotropy effects on flux pinning and flux flow are strongly effective in cuprate as well as iron-based superconductors due to their intrinsically layered crystallographic structure. However $\textrm{Fe(Se,Te)}$  thin films grown on $\textrm{CaF}_2$ substrate result less anisotropic with respect to all the other iron based superconductors. We present the first study on the angular dependence of the flux flow instability, which occurs in the flux flow regime as a current driven transition to the normal state at the instability point ($I^*$,$V^*$) in the current-voltage characteristics. The voltage jumps are systematically investigated as a function of the temperature, the external magnetic field, and the angle between the field and the $\textrm{Fe(Se,Te)}$ film. The scaling procedure based on the anisotropic Ginzburg-Landau approach is successfully applied to the observed angular dependence of the critical voltage $V^*$. Anyway, we find out that $\textrm{Fe(Se,Te)}$ represents the case study of a layered material characterized by a weak anisotropy of its static superconducting properties, but with an increased anisotropy in its vortex dynamics due to the predominant perpendicular component of the external applied magnetic field. Indeed, $I^*$ shows less sensitivity to angle variations, thus being promising for high field applications.
\end{abstract}
\begin{document}

\flushbottom
\maketitle

\thispagestyle{empty}

\section*{Introduction}
The angular dependence of the critical current density $J_c$, the upper critical magnetic field $H_{c2}$ and the irreversibility field $H_{irr}$ as a function of the orientation of the external applied field has been far and wide investigated in High Temperature Superconductors (HTS)\cite{Blatter:1994dg}. A strong influence of anisotropy, layering and finite temperature on the transport properties of these attractive materials for potential applications is well recognized\cite{Blatter:1993dg}. In HTS\cite{Blatter:1994dg} as well as in Iron Based Superconductors (IBS)\cite{Kwok:2016dg}, the goal of increasing $J_c$ and reducing their anisotropy passes through a deep understanding of a complex vortex matter and flux pinning landscapes. A critical issue for applications in high magnetic fields is the current carrying capability of these superconducting materials, which in the IBS case can even be surprising considering similarities and differences between the two classes of superconductors \cite{Tarantini:2016dg}. In particular, the elemental compound $\textrm{Fe(Se,Te)}$ offers several advantages: few non-toxic elements, less complex layered crystal structure, and low anisotropy. Therefore, by assuming $\theta$ the angle between the parallel orientation in the $ab$ planes and the direction of the applied magnetic field, a systematic study of the full current transport $J(\theta,H,T)$ in this material pave the way to set the ultimately limits to the performance of these IBS materials actual competitors of the HTS. 

Historically, the HTS materials are classified as three-dimensional anisotropic or two-dimensional layered superconductors, on the basis that the coherence length in the crystal $c$ direction, $\xi_c(T)$, exceeds the interlayer distance $s$ or they nearly coincide, respectively\cite{Ivlev:1990dg}. It resulted that $\textrm{YBa}_2\textrm{Cu}_3\textrm{O}_{7 \pm \delta}$ (YBCO) belongs to the former class, with an anisotropy parameter $\gamma \le 10$, whereas $\textrm{Bi}_2 \textrm{Sr}_2 \textrm{Ca}_{n-1} \textrm{Cu}_n \textrm{O}_{2n+4+x}$ (BSCCO) to the latter one, with a $\gamma > 10$. As a consequence different types of vortex structures, such as pancake vortices or flux lines, and pinning mechanisms, as the intrinsic or the kink-pinning, have been established depending on the orientation of the applied external field\cite{Feinberg:1990dg}. Furthermore the vortex motion has been investigated by different external driving forces, either thermal or electric\cite{Gray:1991dg}, to get a comprehensive view of flux flow dissipations in these materials.

In the $\textrm{Fe(Se,Te)}$ compound critical currents and its anisotropy have been explored since its discovery\cite{Iida:2013dg}, as well as in other IBS in connection with their multiband nature\cite{Kidszun:2011dg}. Typically the $J_c(\theta,H,T)$ behavior shows a peak corresponding to the field orientations along the $ab$-planes, namely $ab$-parallel peak, and a peak corresponding to the $c$-axis direction, namely $c$-parallel peak; the former due to the intrinsic pinning\cite{Iida:2013dg} or to the presence of planar defects \cite{Civale:2005dg}, whereas the latter is strictly related to an extrinsic pinning mechanism dominated by material defects parallel to the $c$-axis, if present\cite{Eisterer:2011dg}. Moreover, to our knowledge, a direct comparison between intrinsic $ab$-plane pinning and correlated pinning due to planar defects has not been deeply investigated in this $\textrm{Fe(Se,Te)}$ superconductor\cite{Galluzzi:2017dg}, as it has been done in HTS materials\cite{Kwok:1991dg}.
Indeed a different behavior can be observed for the same $\textrm{Fe(Se,Te)}$ compound grown on different substrates\cite{Bellingerisust:2012dg}, by tuning the intrinsic/extrinsic pinning mechanisms. In the search for high-performance high-field superconductors, a good candidate can be the $\textrm{Fe(Se,Te)}$ grown on $\textrm{CaF}_2$ substrate\cite{Weidong:2013dg}, which shows the lowest anisotropy\cite{Braccini:2013dg}, a very robust $J_c(H)$ dependence\cite{Bellingeri:2012dg}, and a sufficient stability against quench under relatively high bias currents\cite{Leo:2016dg,Leo:2017dg}. Additionally, in this compound the study of vortex dynamics has been recently performed at subcritical current values and in self magnetic fields\cite{Nappi:2017dg}. However, in the presence of an external applied magnetic field and at high vortex velocities, we have established the intrinsic nature of a quenching mechanism of the superconducting state, known as flux flow instability, occurring above $J_c$ in the current driven transition to the normal state\cite{Leo:2016dg}. Moreover, several dynamical vortex flow instabilities have been also analyzed in superconductors exhibiting a negative differential resistance\cite{Biplab:2017dg}. Finally, just recently, a sophisticated imaging of ultrafast vortices has been realized at nanometer scale, thus exploring flux flow instabilities by a local probe\cite{Embon:2017dg}, too.

Here we focus on the angular dependence of the flux flow instability, in particular of the critical parameters that are the quenching current $I^*$ and the critical voltage $V^*$, which identifies the upper limits of the flux flow regime, that suddenly is driven into the normal resistive state\cite{Klein:1985dg}. Our purpose is to study the anisotropy of the flux flow instability in an IBS material that is the $\textrm{Fe(Se,Te)}$ compound epitaxially grown on $\textrm{CaF}_2$ substrate (see Methods). On top of that, we evaluate the anisotropy factor $\gamma_J$ defined by the $J_c$ anisotropy, since in multiband materials the anisotropy parameters are different when they are derived from $J_c$, $H_{c2}$ rather than from the penetration depth anisotropy\cite{Kidszun:2011dg}. The low anisotropy values, ranging from 1 to 2, makes this $\textrm{Fe(Se,Te)}$ superconductor different from the other IBS as well as HTS materials, despite the similar layered structure. We find that: ($i$) the critical current $I_c(\theta)$ shows the $ab$-parallel peak at $\theta = 0^\circ$ and $\theta = 180^\circ$ commonly related to the intrinsic layered structure of two-dimensional superconductors, being $\xi_c(T) \leq s$ with $s$ the interlayer distance\cite{Iida:2013dg,Putti:2010dg}, or to correlated planar defects; ($ii$) the observed flux flow critical voltage $V^*$ follows, as a function of the angle $\theta$, the usual scaling of the anisotropic Ginzburg-Landau model\cite{Blatter:1992dg}; ($iii$) surprisingly, the instability current $I^*(\theta)$, typically greater than the critical current, does not follow any available theoretical prediction. In other words, on one hand, there is a layered structure in this material that influences both $V^*$ and $I_c$, but does not affect $I^*$, which gives an higher stability for a better performance in all those applications in which the switching current to the normal state is relevant. On the other hand, experimental data analysis emphasizes a fundamental aspect of the $\textrm{Fe(Se,Te)}$ material similar to some HTS such as BSCCO\cite{Xiao:1999dg}, but it demonstrates a relevant difference on the quenching current of the $\textrm{Fe(Se,Te)}$ compound, which provides more efficient, robust and field independent electric current transport on the orientation of the magnetic field.

\section*{Results}

\begin{figure}[t]
\centering
\includegraphics[width=1\linewidth]{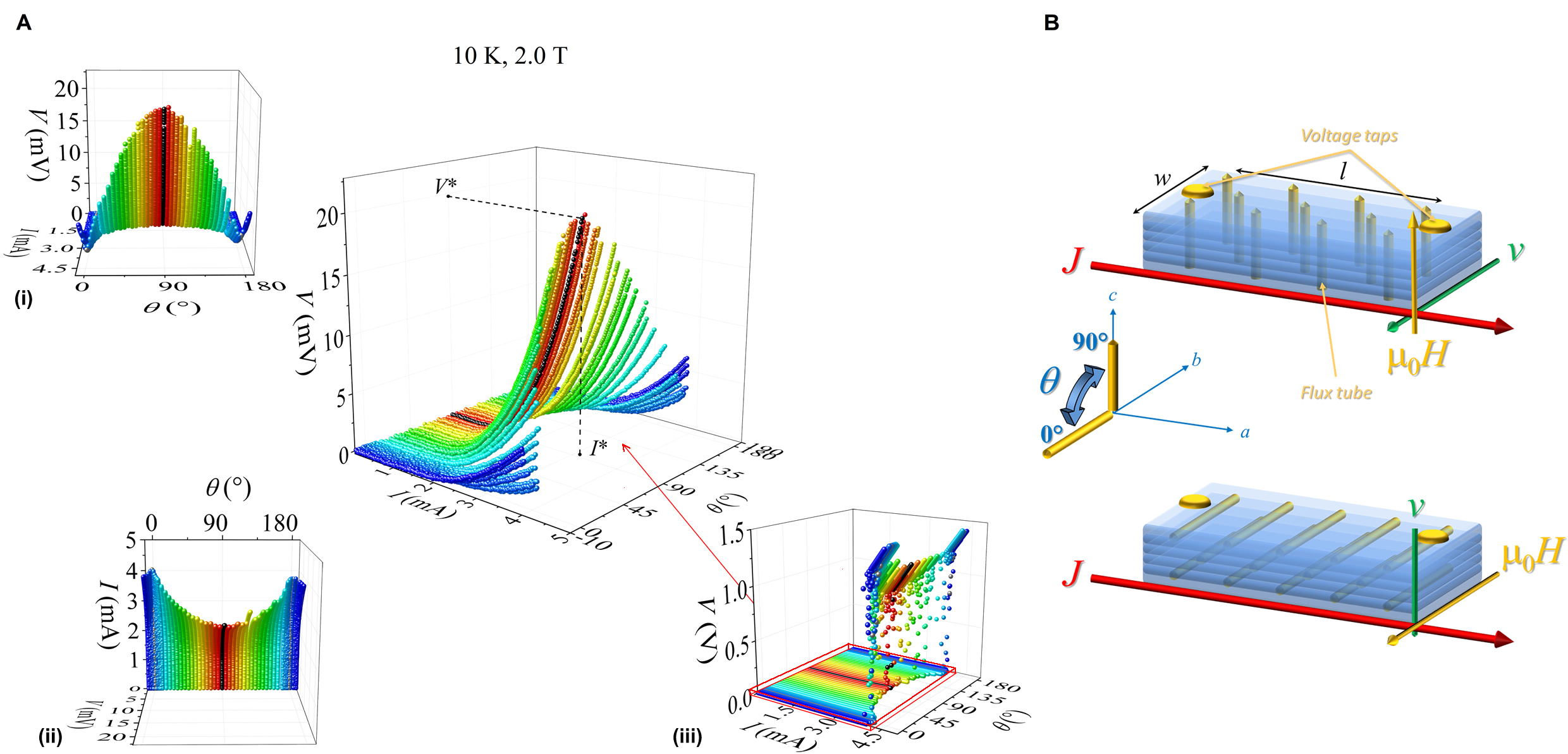}
\caption{A. Current-voltage characteristics at $2$~T and $10$~K as a function of the angle between the applied magnetic field and the direction parallel to the $ab$-planes. The individual angular dependences of $V$ and $I$ are shown (panels ($i$),($ii$)), as well as the ($I$,$V$) curves in the full scale up to the ohmic resistive branches (panel ($iii$)). B. Schematic view of the sample geometry, with the indication of the orientation of the applied magnetic field $H$, the bias current density $J$ and the resulting vortex velocity $v$. The yellow dots represent the voltage taps for voltage measurements as a function of the angle $\theta$.}
\label{fig:1}
\end{figure}

The phenomenon of flux flow instability and its angular dependence is studied as a function of the angle $\theta$ between the external magnetic field and the $ab$-plane at different temperatures and at several values of the applied field. The vertical external magnetic field direction is fixed and the sample is rotated so that the field goes from the direction parallel to $ab$-planes to the direction parallel to the $c$-axis, by always keeping the field perpendicular to the bias current, see Figure~\ref{fig:1}B. Current-voltage $I-V$ measurements are performed by a pulsed current bias technique, in an external field which is provided by an high field superconducting solenoid inside a cryogen-free magnet system (see Methods). Magnetic field-temperature $H-T$ phase diagrams are acquired in the parallel ($\parallel$) and perpendicular ($\perp$) directions in order to extract the material physical parameters useful for the discussion (see Supplemental Materials for details). The microbridges of $\textrm{Fe(Se,Te)}$ material epitaxially grown on $\textrm{CaF}_2$ substrate (see Methods) have the following typical dimensions: thickness $d$ of $150$~nm, width $w$ of $4$~$\mu$m, and length $l$ of $50$~$\mu$m intended as the distance between the voltage taps, as displayed in Figure~\ref{fig:1}B.

The flux flow instability consists in a quenching of the superconducting state, which abruptly drives the system from the dynamic non-linear flux flow resistive state into the normal resistance branch. Its fingerprint is essentially a voltage jump, marked by the critical parameters ($I^*$,$V^*$) in the current-voltage $I-V$ characteristics, see Figure 1A. This corresponds to a current driven transition, which may occur when a sufficiently high bias current destructively perturbs the flux motion out of equilibrium  (see Supplemental Material for more details). 

This phenomenon has been extensively studied in low temperature superconductors (LTS)\cite{Klein:1985dg,Peroz:2005dg,Leo:2011dg,Liang:2010dg,Grimaldi:2009dg}, as well as HTS\cite{Doettinger:1994dg,Kalisky:2006dg,Maza:2008dg}, and very recently in IBS\cite{Leo:2016dg,Leo:2017dg}, too. Its angular dependence has previously been investigated in the BSCCO compound\cite{Xiao:1999dg}, in which strong anisotropy is expected owing to its highly layered structure, i.e. $\xi_c(T) < s$; while for YBCO the study focused on the in-plane angular dependence in connection with the symmetry of the order parameter\cite{Kalisky:2006dg}.

Our $\textrm{Fe(Se,Te)}$ superconducting thin films grown on $\textrm{CaF}_2$ substrates show different pinning contributions: one is anisotropic due to its layered structure or to possible planar defects although not revealed by TEM analysis \cite{Braccini:2013dg}. Another pinning mechanism is ascribed to point-like randomly distributed isotropic defects, which is dominant at the lowest measured temperature of $4.2$ K, that is induced by local modulation of $\textrm{Se}$ and $\textrm{Te}$ stoichiometry, with no $c$-axis correlated pinning\cite{Braccini:2013dg}. Therefore, at increasing temperatures our $\textrm{Fe(Se,Te)}$ films usually show an increasing anisotropy due to pinning. In fact, the critical current behavior becomes more and more anisotropic showing the $ab$-parallel peak more pronounced with respect to the flat angle dependence at lower temperatures \cite{Braccini:2013dg}, as recently even confirmed by other authors \cite{Yuan:2016dg}.

In addition, $\textrm{Fe(Se,Te)}$ can also be considered an highly layered superconductor\cite{Iida:2013dg,Putti:2010dg}, i.e. $\xi_c(T) \leq s$. Therefore, it is intriguing to observe how in dynamic conditions the flux flow instability depends on the angle variation, with a behavior that could be expected similar to BSCCO compound.

\begin{figure}[t]
\centering
\includegraphics[width=0.95\linewidth]{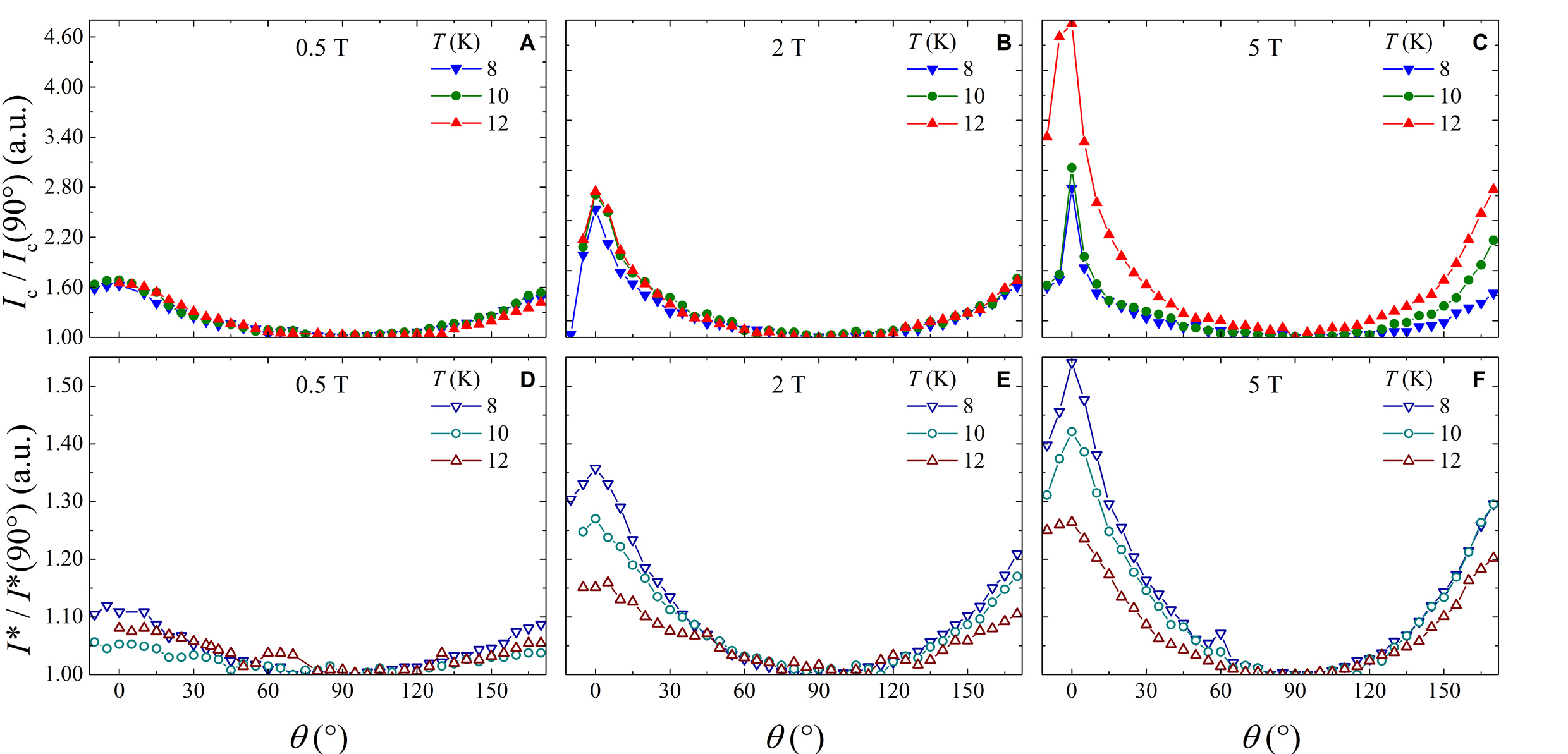}
\caption{The critical current (panels A, B, C) and the instability current (panels D, E, F) as a function of the angle for different field and temperature values reported in the text. Both are normalized to the values corresponding to the field orientation perpendicular to the film surface.}
\label{fig:2}
\end{figure}

\begin{figure}[!hb]
\centering
\includegraphics[width=0.95\linewidth]{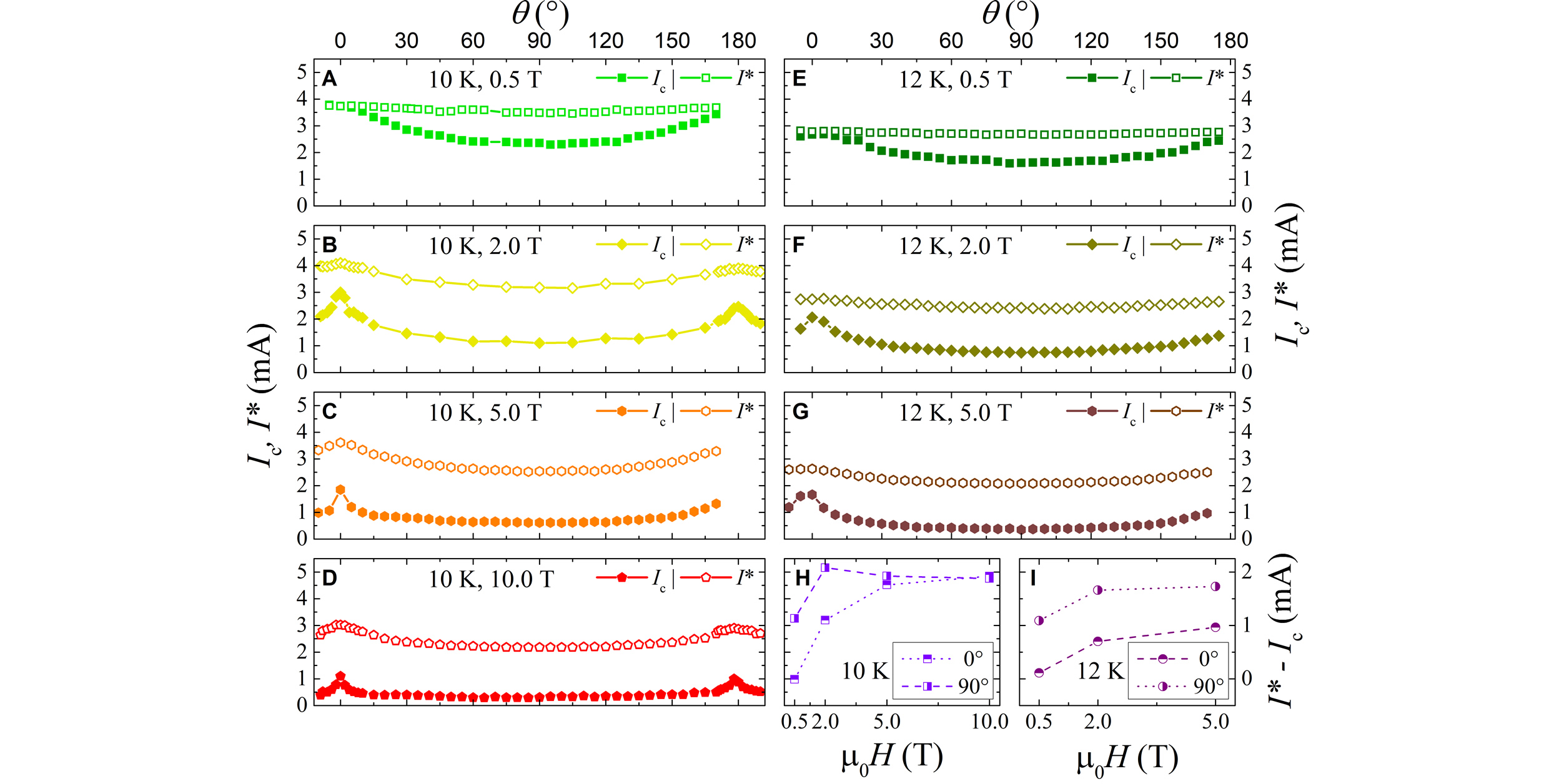}
\caption{The comparison between the critical current and the instability current as a function of the angle for different fields at $10$~K from panel A to D, and at $12$~K from panel E to G. Panels H and I show the difference between the two current values as a function of the applied magnetic field for the two main orientation of the external field and the two measured temperatures of $10$~K and $12$~K.}
\label{fig:3}
\end{figure}

Figure~\ref{fig:1}A shows typical $I-V$ curves as a function of the angle $\theta$ at fixed temperature and at a field intensity value. The comparison of the angular dependence between the critical current and the instability current is reported in Figure~\ref{fig:2} for the three measured temperatures $8$~K, $10$~K, and $12$~K, at different field intensities of $0.5$~T, $2$~T, and $5$~T. The $I_c(\theta)$ behavior reflects the angular dependence usually ascribed to the pinning influence of the layered structure of the $\textrm{Fe(Se,Te)}$ material, as previously found, regardless of the substrate\cite{Iida:2013dg}.The $ab$-parallel peak is expected at $\theta = 0^\circ$ and $\theta = 180^\circ$ and they are actually observed. A scaling versus temperature is observed at all temperatures and low fields by normalizing the $I_c(\theta)$ value to $I_c(90^\circ)$, on the contrary this scaling is lost for the instability current $I^*(\theta)$ at any temperature and field. Moreover, by increasing the magnetic field intensity, the $ab$-parallel peak in the $I_c(\theta)$ becomes more and more pronounced, whereas the $I^*$ does not change so much. In any case no $c$-parallel peak is observed at $\theta = 90^\circ$ that is a consequence of the homogeneous pinning landscape typically induced by the growth on this kind of substrate, as already established\cite{Braccini:2013dg}. By a direct comparison of the $I_c(\theta)$ with the $I^*(\theta)$ values at fixed temperature and different fields in Figure~\ref{fig:3}, it is possible to infer that the instability current is always greater than $I_c$ and in the full measured magnetic field range, the critical current always drops by increasing field  more rapidly than the instability current, whose dependence is quite robust. As a consequence, by increasing the applied magnetic field it results an increase of the difference between the instability and the critical current values, as displayed in Figure~\ref{fig:3} H and I, regardless of the temperature. Anyway, a plateau is reached at very high fields, where a converging behavior is obtained no matter which is the field orientation.

\begin{figure}[!t]
\centering
\includegraphics[width=0.95\linewidth]{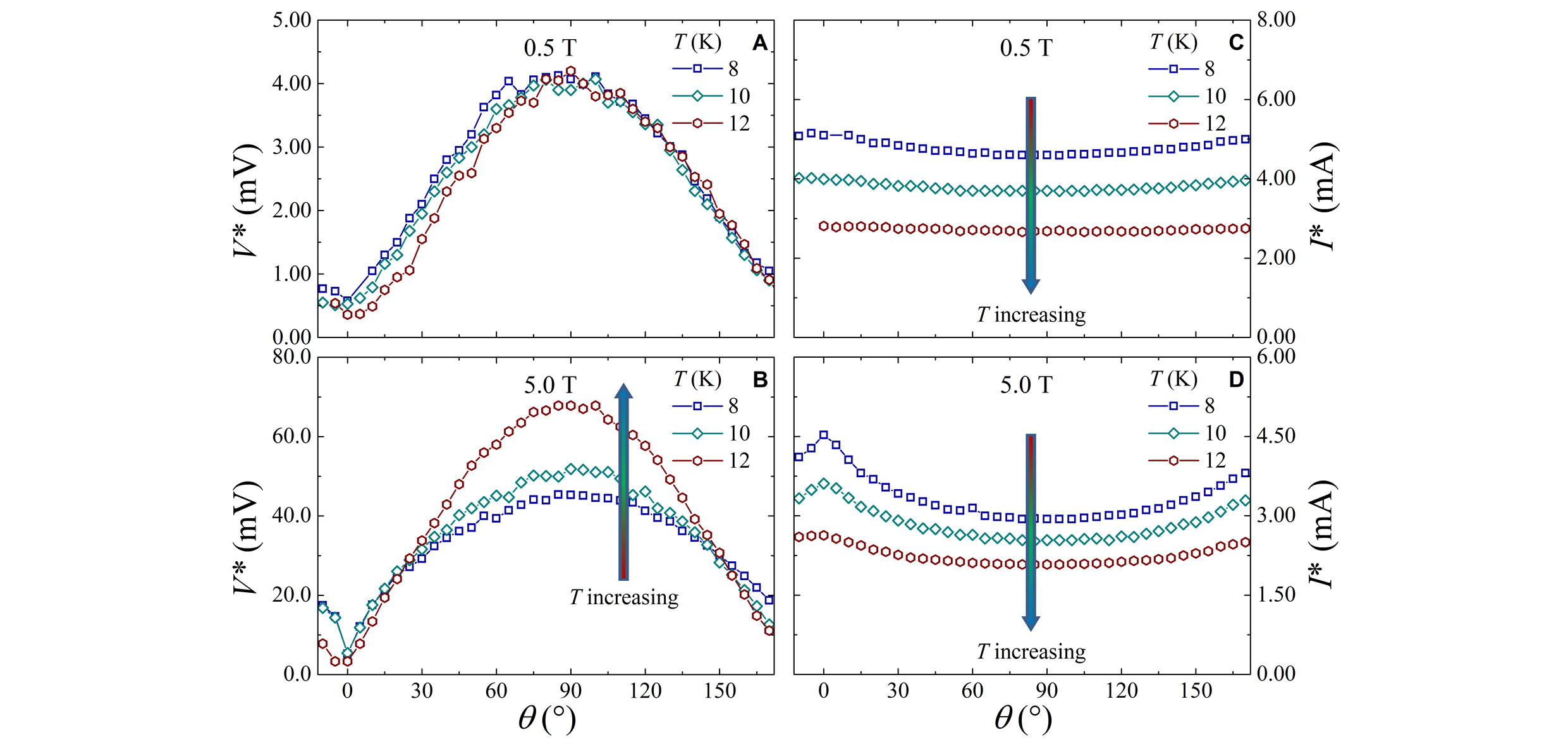}
\caption{The critical parameters of the vortex instability, critical voltage $V^*$ and instability current $I^*$ as a function of the angle for the three measured temperatures of $8$~K, $10$~K, and $12$~K and external field values of $0.5$~T (panels A, C) and $5$~T (panels B, D). The vertical arrows indicates the increasing direction of the temperature.}
\label{fig:4}
\end{figure}

\begin{figure}[!h]
\centering
\includegraphics[width=0.95\linewidth]{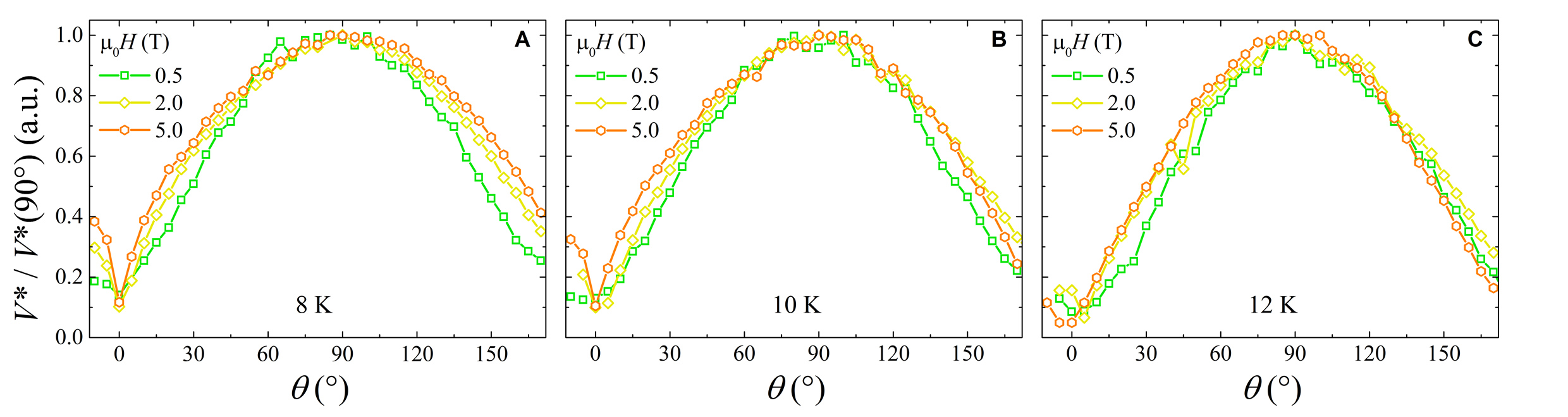}
\caption{The critical voltage as a function of the angle normalized to the values corresponding to the perpendicular direction of the external field. The plots correspond to the full range of measured temperatures and magnetic fields.}
\label{fig:5}

\includegraphics[width=0.95\linewidth]{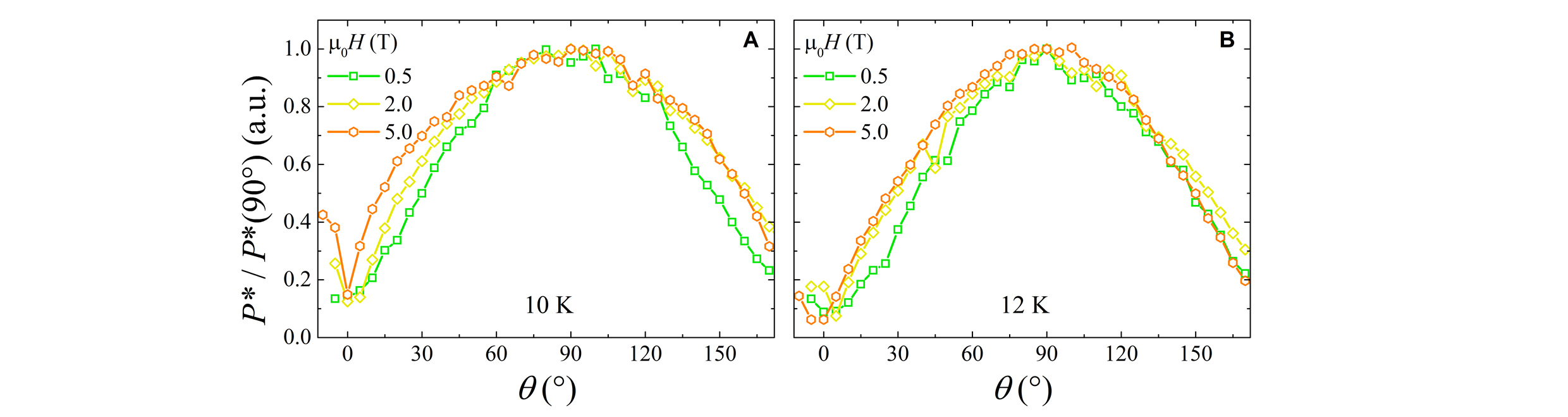}
\caption{The dissipated electric power at the instability point as a function of the angle normalized to the values corresponding to the perpendicular direction of the external field. The plots correspond to the temperatures of $10$~K (panel A) and $12$~K (panel B) in the full magnetic field range investigated.}
\label{fig:6}
\end{figure}

The other critical parameter, which marks the instability of the flux flow state, is the critical voltage $V^*$. The magnetic field and temperature dependence of this voltage is usually investigated in connection with the nature of the intrinsic or extrinsic mechanisms driving the transition to the normal state. In the $\textrm{Fe(Se,Te)}$ superconductor, we have recently demonstrated that a coexistence of both mechanisms is possible\cite{Leo:2016dg}, and that this effect is highly influenced by the microbridge geometry\cite{Leo:2017dg}. In Figure~\ref{fig:4} the $V^*(\theta)$ behavior is displayed at low and high field values, at the three different temperatures $8$~K, $10$~K, $12$~K, in comparison with the corresponding $I^*(\theta)$ dependence. It is worthwhile to remark on the opposite behavior of $V^*$ with respect to $I^*$, since $V^*$ has a maximum at $\theta=90^\circ$ where $I^*$ shows a minimum, even if the former is well pronounced and the latter can be observed only at high fields. We definitely note that at low magnetic field the $I^*$ is much less sensitive to angle variation than $V^*$. In addition, at high field $V^*$ increases as $T$ increases, while $I^*$ decreases. However, by normalizing voltage as $V^*(\theta) / V^*(90^\circ)$, an almost perfect scaling is found at different magnetic fields and temperatures, as displayed in Figure~\ref{fig:5}. Thus, it is clear that the behavior of $V^*(\theta)$ does not follow the same trend of $I^*(\theta)$, and a different scaling rule has to be found in order to interpret this different angle dependences.

Furthermore the angular dependence of the electric dissipated power at the instability point $P^*(\theta) = I^*(\theta) \cdot V^*(\theta)$ is evaluated, with a scaling versus $T$ and $H$ similar to that of $V^*(\theta)$, as shown in Figure~\ref{fig:6}. This physical quantity is usually a measure of the heating effects, since a thermal runway rather than an intrinsic dominant trigger could affect the instability mechanism\cite{Leo:2016dg}. The angular variation of $P^*$ follows strictly the angular variation of $V^*$, by analogy with the BSCCO case\cite{Xiao:1999dg}, but in our case it is mainly due to the blindness of $I^*$ to the same angle variation.

\section*{Discussion}
The impact of flux flow instability on the transport properties of the superconductor has a counterpart on the current driven vortex dynamics at high velocities (see Supplemental Material). Although we are used to correlate flux pinning with critical current, and flux motion with dissipation, by gradually increasing the bias current above $I_c$ the vortex motion can reach a speed limit which is unavoidably affected by pinning mechanism itself\cite{Moll:2013dg}. Either due to material defects\cite{Tarantini:2016dg,Grimaldi:2012dg}, or to geometrical barrier\cite{Papari:2016dg,Grimaldi:2015dg}, pinning becomes crucial in flux flow motion at high velocities too. 
Can the intrinsic layered structure of this $\textrm{Fe(Se,Te)}$ superconductor make feel its influence?

The experimental findings on the several material parameters investigated offer the following scenario: ($i$) $I_c(\theta,H,T)$ shows the only $ab$-parallel peak, which can be interpreted as the fingerprint of the intrinsic layered structure, thus confirming previous results on this $\textrm{Fe(Se,Te)}$ material\cite{Iida:2013dg}. ($ii$) $I^*(\theta,H,T)$ is much less sensitive to angle variation, such that it cannot be formulated within existing theoretical approaches (see Supplemental Material). ($iii$) $V^*(\theta,H,T)$ displays a strong angle dependence with a perfect scaling vs field and temperature, whose interpretation can be found in the framework of anisotropy theories\cite{Blatter:1992dg,Xiao:1999dg}. ($iv$) $P^*(\theta,H,T)$ exhibits a trend similar to $V^*(\theta,H,T)$ that points to the fact that self-heating effects are negligible with respect to electronic ones.\cite{Leo:2016dg} Additionally, the chosen microbridge geometry based on width less than $20$~$\mu$m, as well as the bias current pulsed technique based on $2.5$~ms pulse width, can support the common origin of the voltage jumps in all magnetic field orientations, thus in agreement with our recent study on the same material performed only in perpendicular field configuration\cite{Leo:2017dg,Leo:2016dg}, in which the intrinsic electronic nature of such instability has been demonstrated.

The general problem of anisotropic superconductors has been solved for strong type-II single-band superconductors by the scaling approach of Blatter et al.\cite{Blatter:1992dg}. The scaling rule can be formulated in our case in the following way: $O(\theta,H,T) = s_q O'(\epsilon_{\theta} H,\gamma T)$,being $O$ the observable quantity for which $O'$ is the isotropic corresponding observable quantity, with $s_q = \epsilon$ or $s_q = 1/\epsilon_\theta$ for the observable quantities volume, energy, temperature or magnetic field, respectively.
Here $\epsilon = 1/\gamma$ is the anisotropy factor, and $\epsilon_{\theta}^2 = \epsilon^2 \cos^2 \theta + \sin^2 \theta$ is the scaling factor for magnetic field such that it is $H_{eff} = \epsilon_{\theta} H$.

The anisotropy factor $\epsilon$ can be evaluated by the mass anisotropy ratio $\gamma_m = \left(m_c/m_{ab}\right)^{1/2}$ or by the $J_c$ anisotropy parameter $\gamma_{J}= J_c^{\parallel ab} / J_c^{\parallel c}$, whose magnetic field and temperature dependences can be different since $J_c$ may be affected by factors other than the intrinsic anisotropy\cite{Kidszun:2011dg}. In the strong pinning regime for this multiband superconductor it has already been proven that the scaling procedure for the critical current works by using the temperature dependence of the anisotropy parameter $\gamma_{J}$.\cite{Iida:2013dg} Our estimate gives at lower temperatures an anisotropy value ranging from $1$ to $2$ (see Supplemental Material), which confirms the lower anisotropic pinning character of this material grown on $\textrm{CaF}_2$ substrate\cite{Braccini:2013dg}, more isotropic also with respect to the same superconductor grown on $\textrm{MgO}$ \cite{Iida:2013dg}, as well as other layered 1111-IBS \cite{Tarantini:2016dg}, or 122-IBS \cite{Tarantini:2014dg,Hanisch:2015dg}, too. Therefore in our case, it is possible to apply Blatter scaling rule to verify the anisotropy estimation for $\gamma_{J}$ parameter from the magnetic field dependence of the critical current (see Figure 3 in the Supplemental Material), as usually performed\cite{Iida:2013dg,Yuan:2016dg}.

On the other hand, in dynamic conditions, to describe the angular dependence of the flux flow critical parameters, an analogous approach with this anisotropy scaling could be applied (see Figure 5 in the Supplemental Material), since we observe an evident angular variation in the magnetic field dependence of $V^*$.

\begin{figure}[!t]
\centering
\includegraphics[width=1\linewidth]{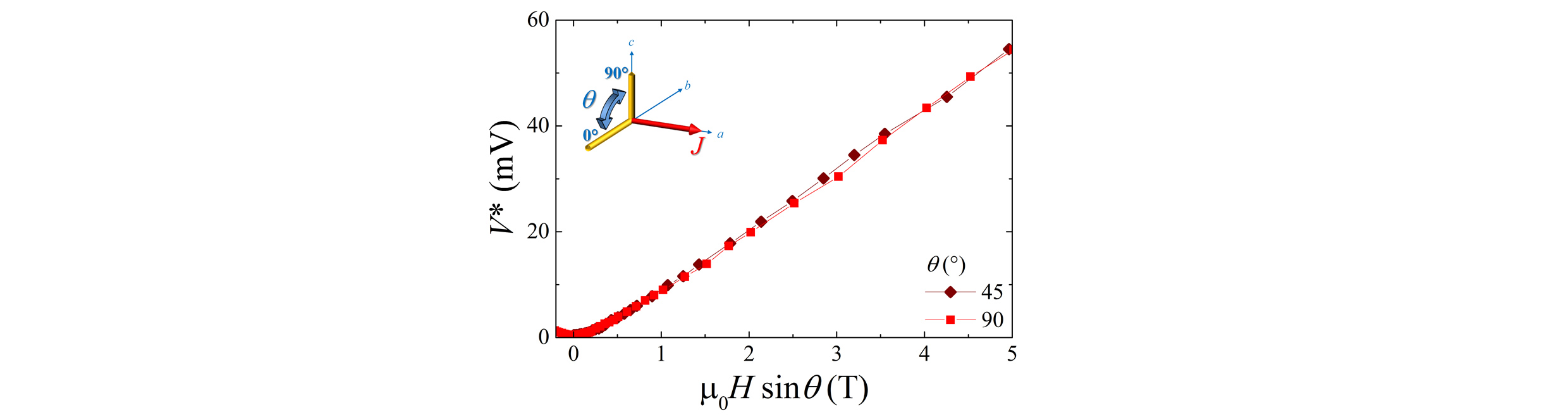}
\caption{The scaling of the critical voltage as a function of the perpendicular component of the applied external magnetic field at $T = 10$~K.}
\label{fig:7}
\end{figure}
 
 Alternatively, it can turn useful the scaling approach followed by Xiao \textit{et al}. in the case of the layered HTS BSCCO compound\cite{Xiao:1999dg}. In fact, in Figure \ref{fig:7} the magnetic field dependence of $V^*$ is plotted at different angles as a function of the field component perpendicular to the $ab$-plane $H \sin \theta$ at $T = 10$~K. Data collapse on a single curve, that is the $V^*$ values are equal providing that $H \sin \theta$ is considered. According to the observed scaling law, it is inferred that $V^*$ measurements are sensitive only to the $c$-axis component of the external applied field $H$, in analogy with the BSCCO case. Nevertheless, this is a further demonstration that Blatter scaling approach can be applied by supposing an higher anisotropy $\gamma \gg 1$ (see Figure 5 in the Supplemental Material). In other words, the limiting behavior for  large $\gamma$ value of the highly anisotropic critical voltage observable $V^*(\theta,H,T)=V^*(\epsilon_{\theta}H,T)=V^*(H \sin \theta,T)$ reconciles both scaling approaches.

On the other hand, this effect can be expected on the general consideration of penetration of vortices in wide superconducting thin films, i.e. $w \gg d$. In tilted applied magnetic fields, it is the perpendicular component that penetrates in form of a vortex lattice, due to large demagnetization factor of thin films \cite{Schmitt:1991dg,Xiao:1999dg}. In a simplified scenario, two types of vortices can be distinguished: vortices that are perpendicular or parallel to the film plane. However, in a layered superconductor it is well known that a static staircase vortex configuration can be established depending on the tilting angle of the external applied magnetic field\cite{Tarantini:2016dg,Kwok:1991dg}. Moreover, our experimental findings for vortex dynamics show that the anisotropy of vortex instability originates from the layering nature, regardless of the geometry effects. Therefore, the dissipation is in practically only due to the component of the applied magnetic field perpendicular to the material layered structure, as in the case of the BSCCO compound extensively studied by Xiao et al.\cite{Xiao:1999dg}.

On the basis of our experimental findings, we can argue that the $\textrm{Fe(Se,Te)}$ is a challenging material that offers some unexpected potentiality, partially reflected in the different observed dependences of $I^*(\theta,H,T)$ with respect to $V^*(\theta,H,T)$. 
Indeed, the two superconductors, BSCCO and $\textrm{Fe(Se,Te)}$, have in common at least the layered crystallographic structure so that they both could show a two-dimensional character, although the anisotropy factor is orders of magnitude different. In any case, this layering character influences the static and dynamics of Abrikosov vortices\cite{Tarantini:2016dg,Moll:2013dg}. Consequently the layering nature of superconductivity is expected to affect the macroscopic quantity $V^*(\theta,H,T)$ in a similar way in the two materials, as demonstrated in Figure \ref{fig:7}. Indeed, there is a direct correlation between the Abrikosov vortex velocity and the measured voltage $V^*$ (see also  Supplemental Material), since by Maxwell equation  $V^* = v^* \cdot \mu_0 H \cdot l$ ($l$ is the distance between voltage pads).

Moreover, the different fabrication processes induce a variety of material defects acting as pinning centers in the two compounds that can also contribute to the different dependences of $I^*(\theta,H,T)$ in the two materials. In fact, the macroscopic quantity of $I^*$ plays the role of the energy supplied to the vortices in the flux flow motion just before the instability takes place, so that any distribution of pinning centers as well as their interactions with the moving vortices can influence such an instability point. In fact, we have previously reported on a tunable pinning strength effect, for example, on $I^*(H)$ behavior\cite{Silhanek:2012dg}. In that case of $Al$ superconducting film with underneath magnetic pinning, we observed a different behavior of $I^*(H)$ and $V^*(H)$. In particular, despite the fact that $I_c$ was affected by pinning influence, $I^*$ resulted practically \textit{insensitive to changes in pinning strength}\cite{Silhanek:2012dg}, but not $V^*(H)$ that followed the expected trend as a function of pinning strength. 
In the absence of a complete theoretical description of flux flow instability, able to take into account not only the material pinning influence but also the multiband peculiar nature of this material, we cannot exclude that the multiband character of $\textrm{Fe(Se,Te)}$, rather different from BSCCO, may influence the electronic mechanism at the base of such instability. Therefore, the present paper may stimulate future works concerning the phenomenon of flux flow instability in IBS materials, since the 11 family already shows remarkable differences with respect to HTS materials, despite their layered similar structures.

By summarizing, the $\textrm{Fe(Se,Te)}$ is characterized by a low anisotropy both in the critical currents and in the magnetic critical fields, although it is a layered superconductor which shows a dynamic behavior much more anisotropic and similar to the cuprates such as the BSCCO material. This result can be interpreted by means of the pinning mechanism acting in the $\textrm{Fe(Se,Te)}$ material. Indeed, at low temperatures and low fields, the pinning centers act isotropically inducing a quite insensitive field dependence of $I_c$. By increasing the temperature and field this anisotropy increases. Above $I_c$ and below $I^*$, from the static staircase configuration of vortices through the layered structure, we pass to a dynamic regime where the dissipation becomes mainly due to the perpendicular component of the applied field. Thus a further increased anisotropic behavior in vortex dynamics is induced. Finally, the instability current $I^*$ also reflects the $ab$-parallel peak symmetry of the layered structure, although it results less sensitive than the critical current.

In conclusion, the $\textrm{Fe(Se,Te)}$ material has some similarities with respect to the HTS materials, but the $I^*$ dependence on the angle variation is very smooth in comparison with that found for for example in BSCCO.  This is indeed another case in which the instability voltage $V^*(\theta, H, T)$ and the instability current $I^*(\theta, H, T)$ are affected by the pinning mechanism \cite{Silhanek:2012dg,Grimaldi:2012dg,Ruck:1997dg}. However, $I^*$ turns to be more robust and isotropic in IBS rather than in HTS, thus promoting this material as an high-field superconductor.

\section*{Methods}
\subsection*{Transport measurements}
Measurements were performed in a Cryogenic, Ltd. Cryogen-Free Measurement System equipped with a variable temperature insert operating in the range $1.6$ to $300$~K with a temperature stability of $0.01$~K, and equipped with a superconducting magnet able to generate up to $16$~T. $I-V$ characteristics were acquired by a pulsed current operation mode specifically developed to minimize self-heating effects. In particular, rectangular pulses of duration $2.5$~ms with an interpulses current-off time of $1$~s was used. The sample temperature has been monitored during each $I-V$ acquisition, which were recorded by increasing and decreasing bias current: no thermal hysteresis has been observed. Data collected as a function of angle variation are obtained by a double-axis rotator probe, on which the sample was mounted with the bias direction always perpendicular to the orientation of the magnetic field. $H-T$ phase diagrams were obtained by $R(T)$ measurements performed with the sample thermally coupled to a copper block in flowing He gas.

\subsection*{Sample fabrication}
Thin films of $\textrm{Fe(Se,Te)}$ epitaxially grown on $\textrm{CaF}_2$ substrates were obtained by pulsed laser deposition from a $\textrm{FeSe}_{0.5}\textrm{Te}_{0.5}$ target with a Nd:YAG laser at $1024$~nm. High quality and purity thin films were deposited at $550 ^\circ$C, on which several microbridges were realized by standard UV photolithography and Ar ion-milling etching. The typical geometry consists of bridges few microns wide, which are all $50$~$\mu$m long. The nominal thickness of the films is $150$~nm. Two identical series of microbridges were obtained on the same thin film.


\section*{Acknowledgements}

The research leading to these results has received funding from the PON Research and Competitiveness 2007-2013 under grant agreement PON NAFASSY, PONa3-00007. The authors wish to thank Y. Bugoslavsky for useful discussions.

\section*{Author contributions statement}

G.G. designed the study, performed experiments and data analysis, wrote the manuscript and planned the figures. A.L. conducted the measurements and prepared the figures. V.B., E.B. and C.F. provided the samples. A.N. and S.P. contributed to the interpretation of the experimental findings. All authors discussed the results and reviewed the manuscript. 

\section*{Additional information}
\textbf{Supplementary information} accompanies this paper at {http://www.nature.com/srep}

\noindent \textbf{Competing financial interests}: The authors declare no competing financial interests.

\newpage
\section*{}
\huge\textbf{\noindent [Supplementary information] Angular dependence of vortex instability in a layered superconductor: the case study of Fe(Se,Te) material}
\normalsize
\input{angular_dep-supplementary_arxiv_v2}

\end{document}

%% file: angular_dep-supplementary_arxiv_v2.tex




\nopagebreak
\setcounter{figure}{0}
\section*{Flux Flow Instability}
The instability of the superconducting state has been studied so far as an interesting phenomenon of vortex dynamics as well as for its relevant impact on the lossless electric current transport in type-II superconductors. No matter is the mechanism triggering the instability, its fingerprint consists of voltage jumps that can be observed in current-driven current-voltage characteristics of the superconducting material. Such jumps can be ascribed to several possible mechanisms, each of which shows its own peculiar feature in the $I-V$ curve branch above the critical current. Here we make a list of the conventional and more exotic ones in connection with their observable fingerprints. \textit{Thermal runway}\citeS{Gurevich:1987dgSM} is well known, since high currents induce a power dissipation in the film that is high enough to destroy the superconducting state, leading to an abrupt increase of sample temperature above $T_c$. \textit{Hot spot effect}\citeS{Gurevich:1987dgSM} is related to a localized normal domain (hot-spot) maintained by Joule heating, usually such domain appears where there is a maximum current concentration; the $I-V$ curve manifests an counterclockwise hysteresis. \textit{Electron overheating}\citeS{Bezuglyj:1992dgSM} is due to the finite heat removal rate of the power dissipated into the sample, depending on the film-substrate interface transparency to phonons, indeed non-equilibrium phonons leave the film without being reabsorbed; therefore the heat removal rate is determined by the strength of the electron-phonon coupling constant rather than by interface properties. \textit{Vortex system crystallization}\citeS{Koshelev:1994dgSM} may occur if the system has enough time to arrange itself into a coherently moving perfect crystal at large velocities; the ordering of the vortex lattice at large applied currents show a jumplike transition between pinned static state and homogeneously moving lattice. \textit{Self-organized criticality}\citeS{Nori:1991dgSM} is marked by voltage instabilities that could appear near the pinning-depinning transition by thermally activated jumps of vortices, which trigger a chain reaction of vortex movements leading to avalanches of diverging size. \textit{Phase-slip centers}\citeS{Ustinov:2003dgSM} (PSC) and/or \textit{lines}\citeS{Vodolazov:2007dgSM} (PSL) occur at currents larger than a certain instability current $I^*$, a system of transverse alternating normal and superconducting domains is formed, and a voltage-step structure in the $I-V$ curve appears; these segments of constant dynamic resistance have a slope independent from the magnetic field strength. PSC appears when the uniform superconducting state is destroyed since the transport current reaches the GL pair-breaking current ($I_c^{GL} < I < I_{c2}$); PSL appears when the steady viscous flux flow of Abrikosov vortices is disrupted at currents $I_m < I_c^{GL}$. The normal state is reached at current higher than the upper critical current $ I > I_{c2} \gg I_m$.

Theoretical approaches within this scenario include the fundamental \textit{theory of Larkin and Ovchinnikov} \citeS{Larkin:1975dgSM} and the \textit{hot electrons instability model} by Kunchur \citeS{kunchur:2002dgSM}. The first predicts the instability from the flux flow regime at temperatures close to $T_c$, caused by the shrinking of the vortex core due to the quasiparticles escaping when a sufficiently high vortex velocity is reached. The second main feature is the expanding of the vortex core as a consequence of the electronic temperature increase which adds quasiparticles within the vortex core at sufficient high electric field but at temperatures far from $T_c$. Nevertheless, both approaches ignores material pinning influence on flux flow instability, since originally they are derived in totally absence of any pinning mechanism. Moreover, just recently, pinning effects have been taken into account in the hot electron flux flow instability by Shklovskij \citeS{Shklovskij:2017dgSM}.  

We have recently demonstrated that some of the aforementioned instability mechanisms may compete depending on the material under investigation, so that in $\textrm{Fe(Se,Te)}$ superconductor the flux flow instability can be considered halfway between those of LTS and HTS, with a coexistence of thermal effects dominated by the electronic nature of the instability\citeS{Leo:2016dgSM}.

\section*{Sample characterization}

\begin{figure}[!t]
\centering
\includegraphics[width=0.95\linewidth]{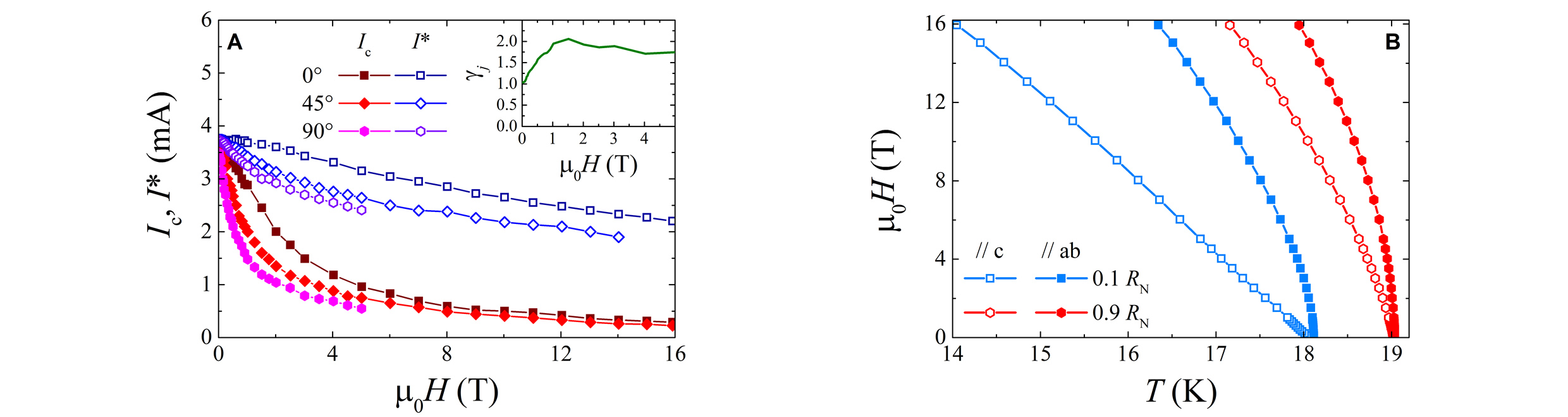}
\caption{A. The critical current and the instability current as a function of the magnetic field intensity for three main orientation of the external field at $T = 10$~K. The inset shows the anisotropy parameter as deduced from the critical current measurements as a function of the field. B. The magnetic field-temperature phase diagram for the two main orientation of the external field.}
\label{fig:SM1}
\end{figure}

\subsection*{Pinning properties}
Transport measurements were performed in order to characterize the pinning properties of this material and their anisotropy. Indeed, the anisotropy factor can be deduced from the critical currents vs field dependence measured in the two field orientation, that is the ratio of $I_c(0^\circ,H)$ to $I_c(90^\circ,H)$. In the Figure \ref{fig:SM1}A the critical current versus $H$ curves are displayed at different magnetic field orientations $\theta = 0^\circ$, $45^\circ$, $90^\circ$, and at fixed temperature $T = 10$~K. The same plot includes the instability currents as well. It is clear that the $I^*$ values are slightly dependent on the intensity and the orientation of the magnetic field. By the way, the $I_c$ values reflects the expected behavior with the parallel in field values always greater than the perpendicular ones. Furthermore, the critical current density is $J_c = 2 \cdot 10^5$~A/cm$^2$ at $10$~K. The inset shows the anisotropy factor $\gamma_J$ versus field dependence, i.e. $I_c(0^\circ)/I_c(90^\circ)$, which results between $1$ and $2$ up to $5$~T. 

\subsection*{Anisotropy properties}
\begin{figure}[!h]
\centering
\includegraphics[width=0.95\linewidth]{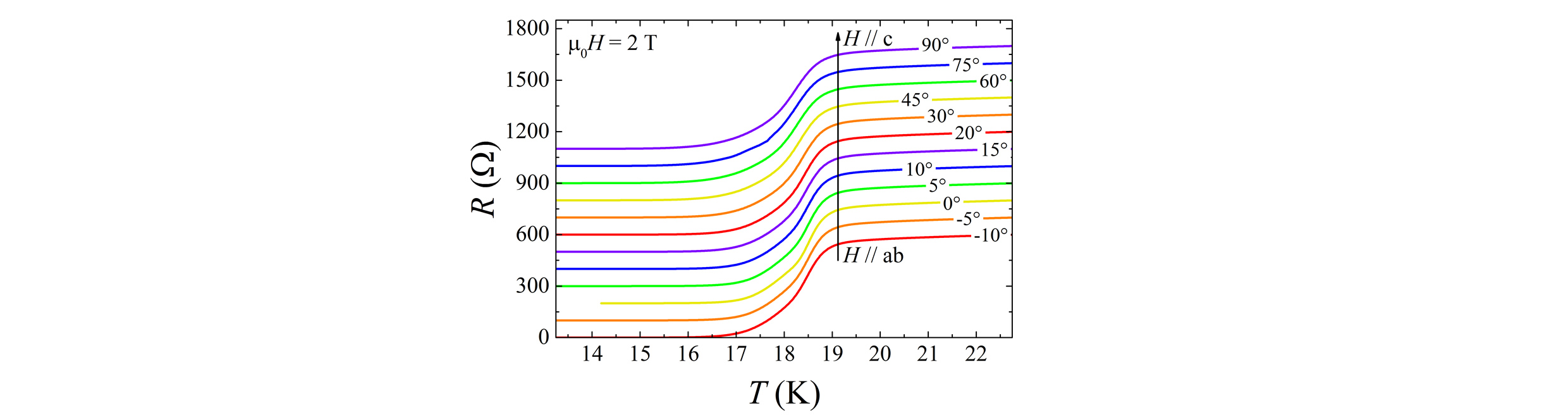}
\caption{The temperature dependent resistance curves at 2~T for different angles from the field parallel to $ab$-planes up to the field parallel to $c$-axis. Curves are shifted by 100~$\Omega$ from each other.}
\label{fig:SM2}
\end{figure}

The high quality of $\textrm{Fe(Se,Te)}$ thin films is also confirmed by the $H-T$ phase diagrams measured in perpendicular and parallel orientation of the applied magnetic field, as reported in Figure \ref{fig:SM1}B. In particular the critical temperature value $T_c$ estimated by the $50\%$ of the normal state resistance in zero field is $18.5$~K, with a transition width below $1$~K. The transition width is defined as the difference between the temperature values corresponding to $90\%$ and $10\%$ of the normal state resistance $R_N$. These are also the two criteria used to identify the upper critical field $H_{c2}$ and the irreversibility lines, respectively, in the phase diagrams shown in Figure \ref{fig:SM1}B. In Figure \ref{fig:SM2} the temperature dependence of the resistance is displayed at different angles from $\theta = 0^\circ$ to $\theta = 90^\circ$ with a field intensity of 2~T.

\begin{figure}[!h]
\centering
\includegraphics[width=0.95\linewidth]{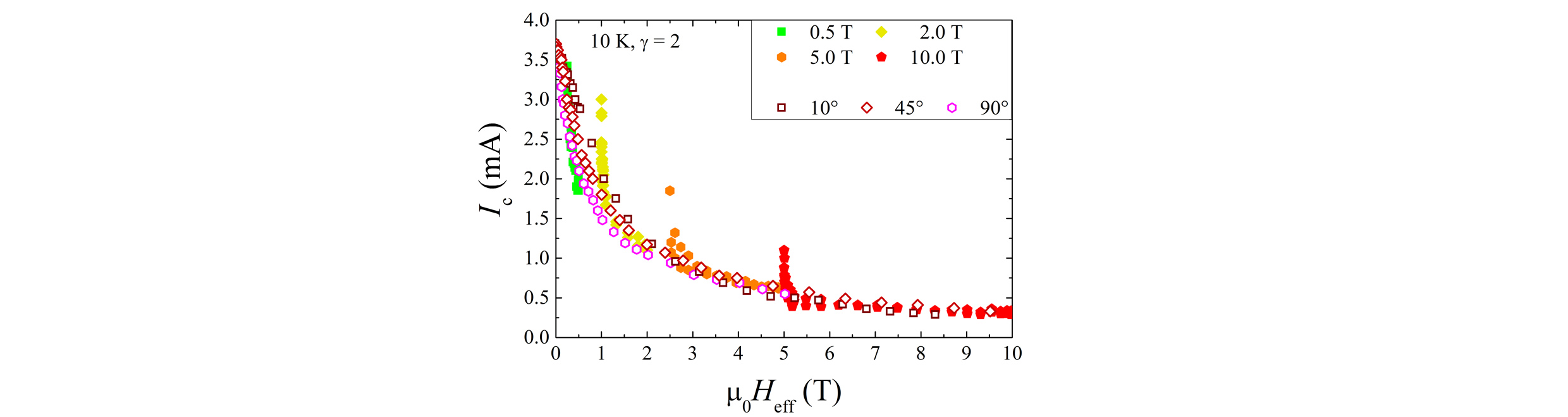}
\caption{The Blatter's scaling curves of the critical currents with $\gamma=2$ for different magnetic field intensities and for different field orientations at $T = 10$~K.}
\label{fig:SM3}
\end{figure}

\begin{figure}[!h]
\centering
\includegraphics[width=0.95\linewidth]{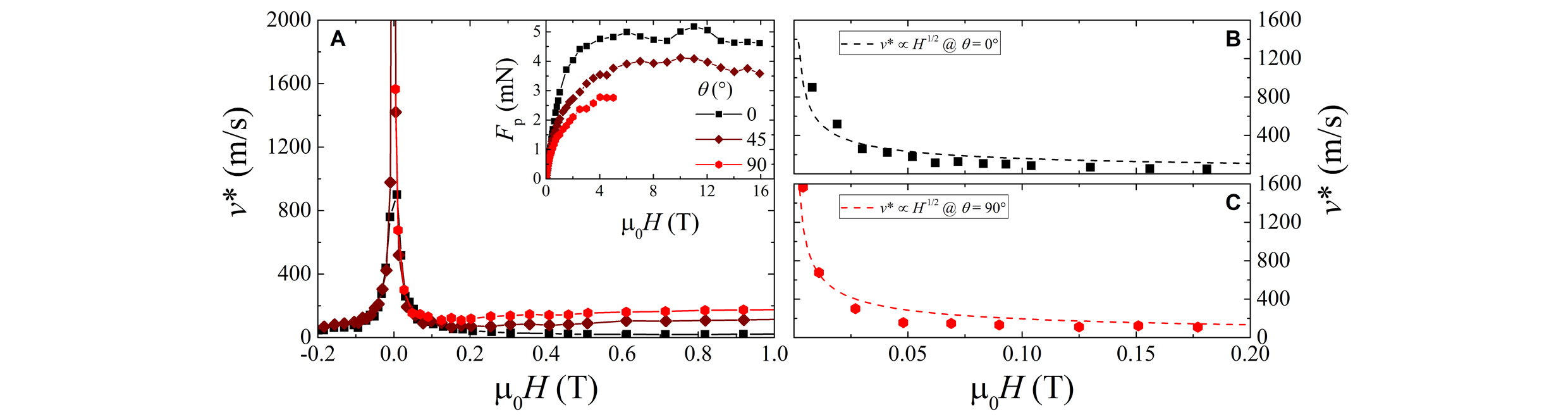}
\caption{A. The vortex critical velocity as a function of the magnetic field intensity for three orientation of the external field at fixed temperature $T = 10$~K. The inset of panel A reports the corresponding pinning force values. B. shows the predicted typical trend (dashed lines) compared with experimental data at the two main orientation of the external magnetic field.}
\label{fig:SM4}
\includegraphics[width=0.95\linewidth]{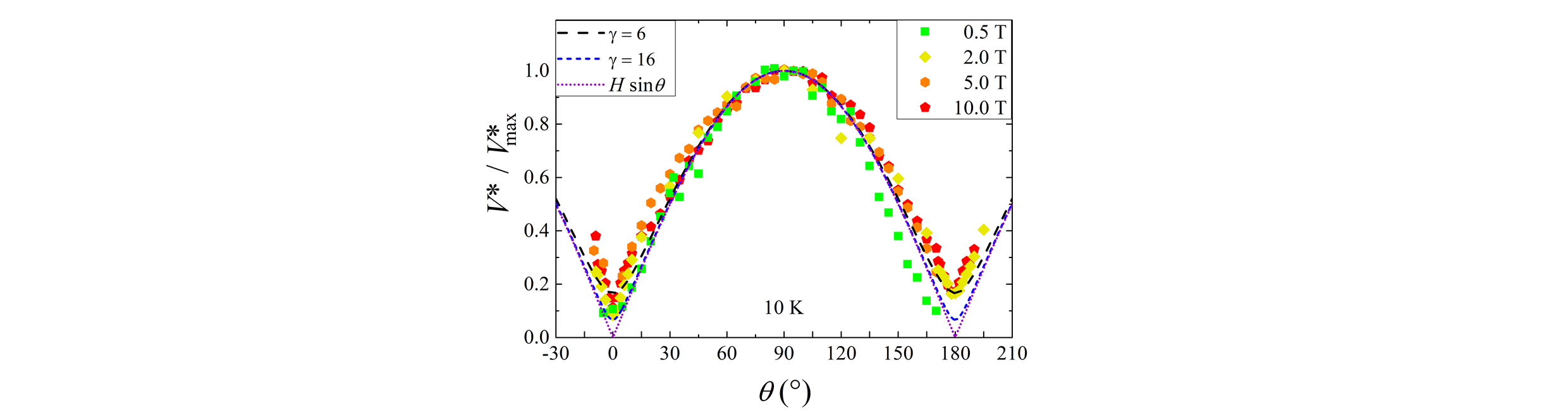}
\caption{The critical voltage as a function of the orientation of the applied magnetic field at fixed temperature $T = 10$~K. The dashed lines show the predicted typical trends following Blatter's scaling approach with a different value of the anisotropy parameter $\gamma$, compared with experimental data at different intensities of the external magnetic field.}
\label{fig:SM5}
\end{figure}

\subsection*{Vortex critical velocity}
By rotating the external field from $\theta = 90^\circ$ to $\theta = 0^\circ$, the vortex velocity results from an in-plane vector velocity towards an out-of-plane vector velocity. In both cases, the resulting longitudinal electric field is detected from the two yellow dots used as voltage taps, as displayed in Figure~1B of the main text. The average vortex critical velocity is given by the measured critical voltage $V^* = v^* \cdot \mu_0 H \cdot l$. In Figure~\ref{fig:SM4}A, the $v^*(H)$ behavior is presented at different values of field orientation equals to $0^\circ$, $45^\circ$, $90^\circ$, and at the fixed temperature of $10$~K. Clearly, in Figure~\ref{fig:SM4}B and C, the typical dependence of $v^* \propto H^{-1/2}$ is observed, which is the expected behavior of the intrinsic electronic nature of flux flow instability\citeS{Doettinger:1994dgSM}. Moreover, it results that $v^*$ increases from parallel (i.e. $\theta = 0^\circ$) to perpendicular (i.e. $\theta = 90^\circ$) orientation of the applied magnetic field. Consequently, in this latter case vortex lattice can move faster. In the inset of Figure~\ref{fig:SM4}A the corresponding pinning force dependence is also shown in order to enlighten that the stronger pinning results when the field is applied in the parallel direction, in agreement with the magnetic field dependence of the critical current (see Figure \ref{fig:SM1}A). In addition we note that there is an increase of $v^*$ vs $H$ at increasing field intensity ($H > 0.1$~T) and for increasing angle orientation ($> 0^\circ$). This is not the only case in which such unusual behavior can be observed\citeS{Dobrovolskiy:2017dgSM}. Probably, this feature can be explained regardless of the specific material under investigation in a more general picture\citeS{Shklovskij:2017dgSM}, but this deserves a more systematic study.  

\bibliographystyleS{plain}
\bibliographyS{bibliographyfile}

